\documentclass[useAMS,usenatbib]{mn2e}
\usepackage{graphicx}
\usepackage{caption}
\usepackage{subcaption}
\usepackage{amssymb}

\def \aap {A\&A} 

\def \apj {ApJ} 
\def \apjl {ApJ} 
\def \apjs {ApJS} 

\def \ga {Geomagn. Aehron.}

\def \grl {Geophys. Res. Lett.}

\def \icarus {Icarus}
\def \i3etm {IEEE Trans. Magn.}
\def \i3etns {IEEE Trans. Nucl. Sci.}
\def \i3etps {IEEE Trans. Plasma Sci.}

\def \jjap1 {Jpn. J. Appl. Phys., Part 1}
\def \jjap2 {Jpn. J. Appl. Phys., Part 2}

\def \mnras {MNRAS} 

\def \nat {Nat} 

\def \pi3e {Proc. IEEE}

\def \pasp {PASP} 

\def \sc {Sci} 

\def \sf {Spaceflight}

\def \usnos4 {U.S. Nav. Obs., Ser. 4}

\usepackage{times}
\usepackage[unicode,colorlinks,urlcolor=cyan,citecolor=blue,linkcolor=blue]{hyperref}

\title[Variability in the super-Earth 55\,Cnc\,e]{Variability in the super-Earth 55\,Cnc\,e}

\author[Demory et al.]
{Brice-Olivier Demory\thanks{E-mail: bod21@cam.ac.uk}$^{1}$, Michael Gillon$^{2}$, Nikku Madhusudhan$^{3}$ \& Didier Queloz$^{1}$ \\
$^1$Astrophysics Group, Cavendish Laboratory, J.J. Thomson Avenue, Cambridge CB3 0HE, UK. \\
$^2$Institut d'Astrophysique et de G\'eophysique, Universit\'e of Li\`ege, all\'ee du 6 Aout 17, B-4000 Li\`ege, Belgium.\\
$^3$Institute of Astronomy, University of Cambridge, Cambridge CB3 0HA, UK.}
 
\begin{document}

\date{Accepted 2015 September 25. Received 2015 September 23; in original form 2015 March 10}

\maketitle

\label{firstpage}

\begin{abstract}
Considerable progress has been made in recent years in observations of atmospheric signatures of giant exoplanets, but processes in rocky exoplanets remain largely unknown due to major challenges in observing small planets. Numerous efforts to observe spectra of super-Earths, exoplanets with masses of 1-10 Earth masses, have thus far revealed only featureless spectra. In this paper we report a 4-$\sigma$ detection of variability in the dayside thermal emission from the transiting super-Earth 55 Cancri e. Dedicated space-based monitoring of the planet in the mid-infrared over eight eclipses revealed the thermal emission from its dayside atmosphere varying by a factor 3.7 between 2012 and 2013. The amplitude and trend of the variability are not explained by potential influence of star spots or by local thermal or compositional changes in the atmosphere over the short span of the observations. The possibility of large scale surface activity due to strong tidal interactions possibly similar to Io, or the presence of circumstellar/circumplanetary material appear plausible and motivate future long-term monitoring of the planet.
\end{abstract}


\section{Introduction}

\label{intro}

A key question in exoplanet science is about the origin of planets in tight orbit (P$<$0.75 days) around their stars. During its primary mission, {\it Kepler} enabled the detection of a hundred ultra-short period (USP) planet candidates \citep{Sanchis-Ojeda:2014}. The majority of these USP planets are small ($R_P<2 R_{\oplus}$) and may suggest that they have experienced a phase of intense erosion due to tidal activity and/or intense stellar irradiation \citep[e.g.][]{Owen:2013}.

One example of these disintegrating USP planets is KIC12557548b \citep{Rappaport:2012}, which exhibits changes in the optical transit depth and shape that seem consistent with an evolving cometary-like environment \citep{Brogi:2012,Rappaport:2014a,Croll:2014,Bochinski:2015}. Scattering of dust particles surrounding the planet are thought to be the driver of the observed variability, as the engulfed sub-Mercury-size object itself would be too small to be actually detected \citep{Perez-Becker:2013}. The high effective temperatures of USP planets could enable a silicate-rich atmosphere made of refractory material to be sustained, which would shape the overall opacity. Replenishment of grains in the atmosphere could be achieved by condensation at cooler temperatures near the day-night terminator or at lower pressures \citep{Schaefer:2009,Castan:2011}. Another source for the observed refractory material could be volcanism, fuelled by tidal heating \citep{Jackson:2008a,Barnes:2010} similar to Io, where the interactions between an inhomogeneous circumplanetary neutral cloud and a circumstellar charged torus \citep{Belcher:1987} would result in transit depth variability over timescales that are similar to the planet's orbital period.

The super-Earth 55\,Cancri\,e has a mass of $8.09 \pm 0.26\: M_\oplus$ \citep{Nelson:2014a}, a published radius of $2.17 \pm 0.10\: R_\oplus$ \citep{Gillon:2012}, and orbits a nearby sun-like star with an 18-hour period. 55\,Cnc\,e is one of the largest members of known USP planets. Owing to its close proximity to the parent star, 55\,Cnc\,e has an equilibrium temperature of $\sim$2400 K and may experience intense tidal heating \citep{Bolmont:2013}.  With V$\sim$6, 55 Cnc is the brightest star known to host a transiting exoplanet, and has led to measurements of the planet's radius at exquisite precision in the visible as well as in the Spitzer 4.5 $\mu$m IRAC band \citep{Demory:2011,Winn:2011a,Gillon:2012}. Thermal emission measurements in the Spitzer 4.5 $\mu$m IRAC photometric band \citep{Demory:2012} yielded a brightness temperature of about 2,300\,K in this channel. Such elevated temperature has implications on the nature of the planet atmosphere. First, this high temperature means that for a given chemical composition, 55 Cnc e would have a larger atmospheric scale height compared to a cooler planet with the same composition. Second, clouds are less likely to form in the atmospheres of extremely hot exoplanets compared to cooler objects \citep{Madhusudhan:2015}, which is particularly relevant in the light of the numerous non-detections of spectral features in exoplanets \citep[e.g.][]{Pont:2008,Sing:2011b,Knutson:2014a,Kreidberg:2014}. The combination of the high photometric precision possible for transit/occultation observations and the high planetary temperature means that 55 Cnc e currently represents one of the most promising super-Earth for infrared spectroscopic observations of its atmosphere. Using the published mass and radius alone, the interior composition of the planet is consistent with: (1) a silicate-rich interior with a dense H$_2$O envelope of 20\% by mass \citep{Demory:2011}, (2) a purely silicate planet with no envelope \citep[e.g.][]{Gillon:2012}, or (3) a carbon-rich planet with no envelope \citep{Madhusudhan:2012a} among other compositions.

In this paper, we present an analysis of archival and new {\it Spitzer}/IRAC data obtained in 2012 and 2013. This combined dataset suggests a radically different nature for 55\,Cnc\,e that could be more consistent with a scenario similar to KIC12557548b than the more conventional picture that was attached to 55\,Cnc\,e and which is described above. We present the observations and data analysis in Section~2, a comparison with previous results in Section~3 and we finally propose an interpretation of our results in Section~4.

\section{Data analysis}

\subsection{Observations}

This study is based on six transits and eight occultations of the super-Earth 55\,Cnc\,e acquired between 2011 and 2013 in the 4.5-$\mu$m channel of the {\it Spitzer Space Telescope} \citep{Werner:2004} Infrared Array Camera (IRAC) \citep{Fazio:2004a}. Data were obtained as part of the programs IDs 60027, 80231 and 90208. Two transits were observed in 2011 (PID 60027) \citep{Demory:2011,Gillon:2012}, four occultations in 2012 (PID 80231) \citep{Demory:2012}, as well as four transits and four other occultations in 2013 (PID 90208). The corresponding data can be accessed using the Spitzer Heritage Archive \footnote{Spitzer Heritage Archive: http://sha.ipac.caltech.edu}. All {\it Spitzer} observations have been obtained with the same exposure time (0.02 s) and observing strategy (stare mode). The last three occultations of PID 80231 have been obtained using the PCRS peak-up mode as all of our 2013 data. The PCRS peak-up procedure is designed to place the target on a precise location on the detector to mitigate the intra-pixel sensitivity \citep[][and references therein]{Ingalls:2012}. The AORs 48072704 and 48072960 experienced 30-min long interruptions that happened during the ingress and egress of 55\,Cnc\,e's transit respectively.

\subsection{Data reduction}

The starting point of our reduction consists of the basic calibrated data (BCD) that are FITS data cubes encompassing 64 frames of 32x32 pixels each. Our code sequentially reads each frame, converts fluxes from the {\it Spitzer} units of specific intensity (MJy/sr) to photon counts, and transforms the data timestamps from BJD$_{\rm UTC}$ to BJD$_{\rm TDB}$ \citep{Eastman:2010a}. The position on the detector of the point response function (PRF) is determined by fitting a Gaussian to the marginal X, Y distributions using the {\sl GCNTRD} procedure part of the IDL Astronomy User's Library \footnote{IDL Astronomy User's Library: http://idlastro.gsfc.nasa.gov}. We also fit a two-dimensional Gaussian to the stellar PRF \citep{Agol:2010}. We find that determining the centroid position using {\sl GCNTRD} results in a smaller dispersion of the fitted residuals by 10 to 15\% across our dataset, in agreement with other warm {\it Spitzer} analyses \citep{Beerer:2011}. We then perform aperture photometry for each dataset using a modified version of the {\sl APER} procedure using aperture sizes ranging from 2.2 to 4.4 pixels in 0.2 pixels intervals. The background apertures are located 10 to 14 pixels from the pre-determined centroid position. We also measure the PRF full width at half maximum along the X and Y axes. We use a moving average based on forty consecutive frames to discard datapoints that are discrepant by more than 5$\sigma$ in background level, X-Y position and FWHM. In average, we find that 0.5\% of the datapoints acquired without the PCRS peak-up mode are discarded, while it is only 0.06\% when the PCRS peak-up mode is enabled. The resulting time-series are finally binned per 30 s to speed up the analysis. The product of the data reduction consists of a photometric datafile for each aperture (12 in total per dataset). We then use the out-of-eclipse portion of the light curve to measure the RMS and level of correlated noise (see Sect.~\ref{beta}) to determine the aperture that minimises both quantities. In the cases of AORs 48070144, 48073472 and 48073728 we found that two to three aperture sizes produced similarly small RMS and beta factors. We selected the one with the smallest RMS. For these three AORs, repeating the analysis using the other aperture sizes does not change the retrieved occultation depths. We show the retained aperture size as well as the corresponding RMS and $\beta$ factor for each dataset in Table~\ref{tab:obs}. For all datasets we discard the photometry corresponding to the first 30 minutes as it usually exhibits a strong time-dependent ramp \citep{Demory:2011}. This procedure is common practice for warm {\it Spitzer} data \citep[e.g.][]{Deming:2011a}.

\begin{table*}
\begin{tabular}{llllllll}

\hline
Date [UT] & Program ID & AOR \# & AOR duration [h] & Eclipse type & Aperture [pix] & RMS/30s [ppm] & $\beta_r$ \\
\hline

2011-01-06 & 60027 & 39524608 & 4.9 &Transit & 2.6 & 359 & 1.75 \\
2011-06-20 & 60027 & 42000384 & 5.9 & Transit & 3.2 & 367 & 1.45 \\
2012-01-18 & 80231 & 43981056 & 5.9 & Occultation & 2.8 & 364 & 1.54 \\
2012-01-21 & 80231 & 43981312 & 5.9 & Occultation & 3.2 & 329 & 1.04 \\
2012-01-23 & 80231 & 43981568 & 5.9 & Occultation & 3.2 & 337 & 1.00 \\
2012-01-31 & 80231 & 43981824 & 5.9 & Occultation & 3.2 & 362 & 1.27 \\
2013-06-15 & 90208 & 48070144 & 5.9 & Occultation & 2.6 & 323 & 1.12 \\
2013-06-21 & 90208 & 48070656 & 8.8 & Transit & 2.8 & 365 & 1.88 \\
2013-07-03 & 90208 & 48072448 & 8.8 & Transit & 3.2 & 350 & 1.16 \\
2013-07-08 & 90208 & 48072704 & 8.8 & Transit & 2.6 & 378 & 1.37 \\
2013-07-11 & 90208 & 48072960 & 8.8 & Transit & 2.6 & 400 & 1.85 \\
2013-06-18 & 90208 & 48073216 & 8.8 & Occultation & 3.0 & 325 & 1.59 \\
2013-06-29 & 90208 & 48073472 & 8.8 & Occultation & 3.0 & 357 & 1.33 \\
2013-07-15 & 90208 & 48073728 & 8.8 & Occultation & 3.4 & 381 & 1.25 \\

\end{tabular} 
\caption{\label{tab:obs} {\bf 55\,Cnc\,e Spitzer dataset.} Astronomical Observation Request (AOR) properties for 2011-2013 Spitzer/IRAC 4.5-$\mu$m data used in the present study.}

\end{table*}

\subsection{Photometric Analysis}

We split the analysis in three steps. In a first step we focus on the modelling of the intra-pixel sensitivity and the photometric precision. The second step focuses on an improved determination of the system parameters using six transits and eight occultations. The third step details the analysis focusing on the variations in transit and occultation depths.

\subsubsection{Intra-pixel sensitivity correction}
\label{ips}
Warm Spitzer data are affected by a strong intra-pixel sensitivity that translates into a dependence of the stellar centroid position on the pixel with the measured stellar flux at the 2\% level \citep{Charbonneau:2005}. A usual method is to detrend the measured flux from the X/Y position of the centroid, using quadratic order polynomial functions. In the past years, several studies \citep{Ballard:2010b,Stevenson:2012a,Lewis:2013,Deming:2014} have employed novel techniques that improved the mitigation of the intra-pixel sensitivity in {\it Spitzer}/IRAC 3.6 and 4.5$\mu$m channels. 

It has been shown that in some cases, the technique used to correct the intra-pixel sensitivity could have a significant impact on the retrieved system parameters in general and the eclipse depth in particular \citep{Deming:2014}. In the present work, we use a modified implementation of the BLISS (BiLinearly-Interpolated Sub-pixel Sensitivity) method \citep{Stevenson:2012a}.

The BLISS algorithm uses a bilinear interpolation of the measured fluxes to build a pixel-sensitivity map. The data are thus self-calibrated. We include this algorithm in the Markov Chain Monte Carlo (MCMC) implementation already presented in the literature \citep{Gillon:2012a}. The improvement brought by any pixel-mapping technique such as BLISS requires that the stellar centroid remains in a relatively confined area on the detector, which warrants an efficient sampling of the X/Y region, thus an accurate pixel map. In our implementation of the method, we build a sub-pixel mesh made of $n^2$ grid points, evenly distributed along the $x$ and $y$ axes. The BLISS algorithm is applied at each step of the MCMC fit. The number of grid points is determined at the beginning of the MCMC by ensuring that at least 5 valid photometric measurements are located in each mesh box. Similar to \citet{Lanotte:2014}, we find that the PRF's FWHM along the $x$ and $y$ axes evolve with time and allows further improvement on the systematics correction. We thus combine the BLISS algorithm to a linear function of the FWHM along each axis. We find that including a model of the FWHM decreases the Bayesian Information Criterion \citep[BIC,][]{Schwarz:1978}  for all datasets. Including a linear dependence with time does not improve the fit. We find no correlation between the photometric time-series and the background level. We finally perform the same intra-pixel sensitivity correction using a variable aperture in the reduction and including the noise pixel \citep{Lewis:2013} parameter as an additional term in the parametric detrending function. As noticed in previous studies \citep{Lewis:2013}, we find this approach slightly increases the BIC in the 4.5$\mu$m channel. We repeat this part of the analysis with lightcurves using apertures sizes smaller and larger by 0.2 pixels from the one minimising the photometric RMS. We subtract the lightcurves using the altered aperture sizes to the optimal one and find point-to-point variations embedded in the 1-$\sigma$ individual error bars. We further check the incidence of binning on the efficiency of the intra-pixel sensitivity correction by analysing the 2012-01-18 occultation (obtained with the PCRS mode enabled) with binnings ranging from 0.64s (one integrated subarray data cube), 6s, 12s and 30s. We find the following depths: 106$\pm$55, 91$\pm$57, 111$\pm$60 and 87$\pm$53 ppm respectively. We thus conclude that using binnings up to 30s does not affect the occultation depth measurement.

\subsubsection{Noise budget and photometric precision}
\label{beta}
The data reduction described above computes the theoretical error for each photometric datapoint. In order to derive accurate uncertainties on the parameters, we evaluate the level of correlated noise in the data following a time-averaging technique detailed in \citet{Pont:2006b}. We compute within the MCMC fit a correction factor $\beta$ based on the standard deviation of the binned residuals for each light curve using different time-bins. We keep the largest correction factor $\beta$ found for each dataset and multiply the theoretical errors by this factor. We show on Table~\ref{tab:obs} the photometric RMS and $\beta$ for each dataset. We also conduct a residual permutation bootstrap analysis to obtain an additional estimation of the residual correlated noise. We use for this purpose the lightcurve corrected from the systematic effects described in Sect.~\ref{ips}. The resulting parameters are in agreement with the ones derived from the MCMC analysis (Table~\ref{tab:res}), while their error bars are significantly smaller. This result indicates that the error budget is dominated by the baseline model uncertainties and not by the residual correlated noise.

\subsubsection{Refinement of the system parameters}
\label{syspar}

In the second step of our analysis, we use as input data to the MCMC all six transits and eight occultations. The goal of this part of the analysis is to refine the published parameters for the orbital and physical parameters of 55\,Cnc\,e. The jump parameters are the planetary-to-star radius ratio $R_{P}/R_{\star}$, the planetary impact parameter $b$, orbital period $P$, transit duration $W$ and centre $T_0$, and the occultation depth $\delta_{occ}$. In a first MCMC run, we keep the eccentricity fixed to zero as previous studies using radial-velocity data, which provides better constraints on $\sqrt{e} \sin \omega$, found eccentricity values consistent with 0 \citep{Nelson:2014a}. We also add the limb-darkening linear combinations $c_1=2u_1+u_2$ and $c_2=u_1-2u_2$ as jump parameters, where $u_1$ and $u_2$ are the quadratic coefficients drawn from the theoretical tables of \citet{Claret:2011} using published stellar parameters \citep{von-Braun:2011a}. We impose Gaussian priors on the limb-darkening coefficients and the radial-velocity semi amplitude $K$ \citep{Nelson:2014a}. We execute two Markov chains of 100,000 steps each and assess their efficient mixing and convergence using the Gelman-Rubin statistic \citep{Gelman:1992} by ensuring $r <1.01$. Results for this MCMC fit are shown on Table~\ref{tab:res}. In a second MCMC run, we explore what limits can be placed on the orbital eccentricity $e$ and argument of periastron $\omega$. We perform a fit with $\sqrt{e}\cos{\omega}$ and $\sqrt{e}\sin{\omega}$ as additional free parameters and find $\sqrt{e}\cos{\omega}=0.01_{-0.04}^{+0.04}$ and $\sqrt{e}\sin{\omega}=-0.05_{-0.17}^{+0.14}$, corresponding to a 3-sigma upper limit on the eccentricity of $e < 0.19$. We do not detect variations in the transit or occultation timings.

\begin{table*}
\begin{tabular}{ll}
\hline
Planet/star area ratio $R_p/R_s$ & $0.0187^{+0.0007}_{-0.0007}$ \\
$b=a \cos i /R_{\star}$ [$R_{\star}$] & $0.36^{+0.07}_{-0.09}$ \\
$T_0 - 2,\!450,\!000$ [BJD$_{\rm TDB}$]& $5733.008^{+0.002}_{-0.002}$ \\
Orbital semi-major axis $a$ [AU]  & $0.01544^{+0.00009}_{-0.00009}$ \\
Orbital inclination $i$ [deg]  & $83^{+2}_{-1}$ \\
Mean density $\rho_{p}$ [g\, cm$^{-3}$]  &$6.3^{+0.8}_{-0.7}$ \\
Surface gravity $\log g_p$ [cgs]  & $3.33^{+0.04}_{-0.04}$ \\
Mass $M_{p}$ [$M_{\oplus}$]  & $8.08^{+0.31}_{-0.31}$ \\
Radius $R_{p}$ [$R_{\oplus}$]  & $1.92^{+0.08}_{-0.08}$ \\
$\delta_{occ}$ (2012) [ppm] & $47\pm21$ \\
$\delta_{occ}$ (2013) [ppm] & $176\pm28$ \\
\end{tabular} 
\caption{\label{tab:res} {\bf 55\,Cnc\,e system parameters.} Results from the MCMC combined fit. Values indicated are the median of the posterior distributions and the 1-$\sigma$ corresponding credible intervals.}
\end{table*}

\subsubsection{Transit depth variations}
\label{supp:tr}

\begin{figure*}
\centering
\includegraphics[scale=0.8]{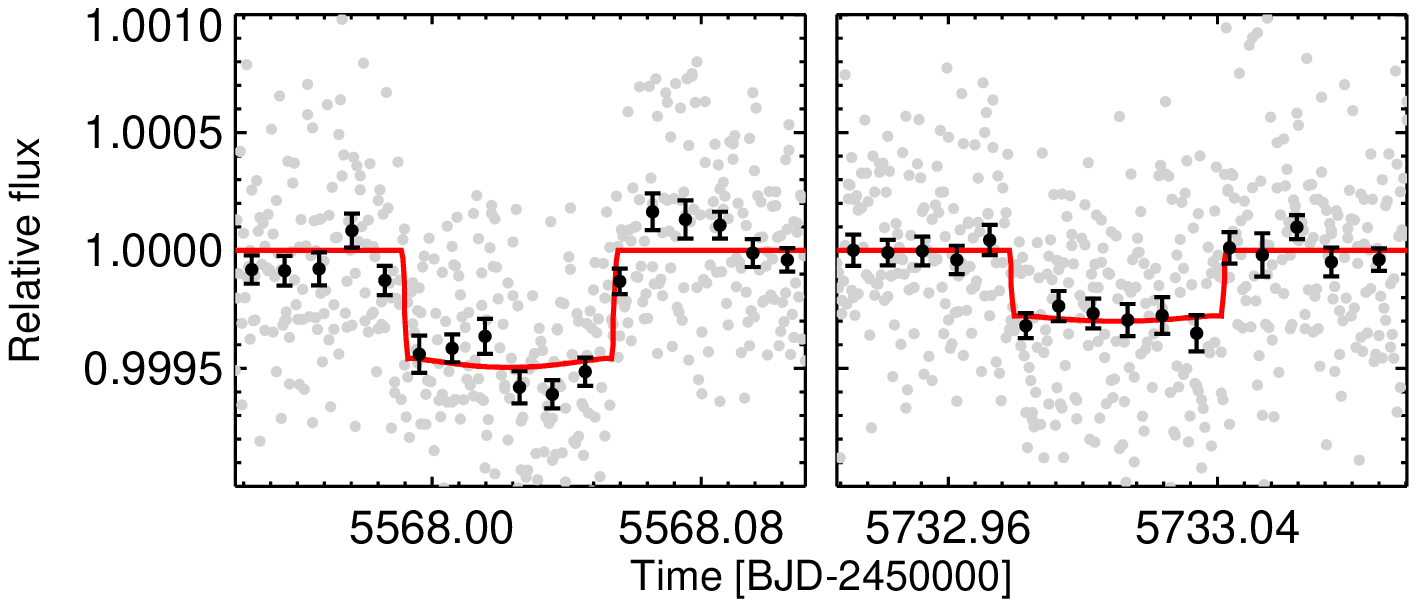}
\caption{\label{fig:tr11}{\bf 2011 transit lightcurves.} Individual analyses of the Spitzer/IRAC 4.5-$\mu$m 2011 transit data detrended and normalised following the technique described in the text. Black filled circles are data binned per 15 minutes. The best-fit model resulting from the MCMC fit is superimposed in red for each eclipse.}
\end{figure*}

\begin{figure*}
\centering
\includegraphics[scale=0.6]{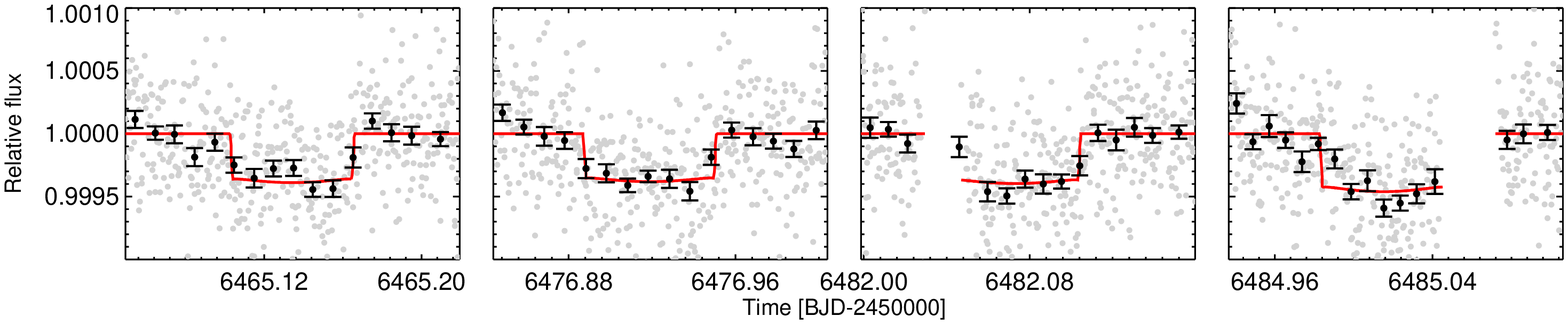}
\caption{\label{fig:tr13}{\bf 2013 transit lightcurves.} Individual analyses of the Spitzer/IRAC 4.5-$\mu$m 2013 transit data detrended and normalised following the technique described in the text. Black filled circles are data binned per 15 minutes. The best-fit model resulting from the MCMC fit is superimposed in red for each eclipse.}
\end{figure*}

\begin{figure*}
\centering
\includegraphics[scale=0.6]{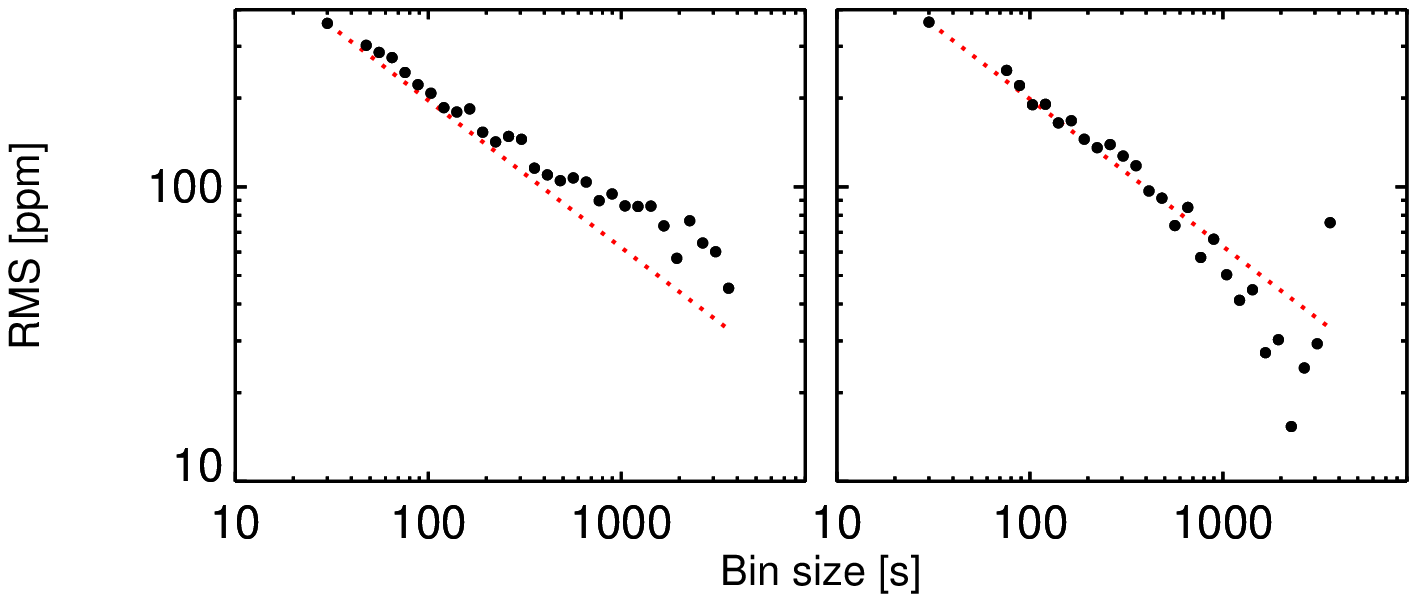}
\caption{\label{fig:trrms11}{\bf 2011 transit  - RMS vs. binning.} Black filled circles indicate the photometric residual RMS for different time bins. Each panel corresponds to each individual transit from 2011. The expected decrease in Poisson noise normalised to an individual bin (30s) precision is shown as a red dotted line. }
\end{figure*}

\begin{figure*}
\centering
\includegraphics[scale=0.6]{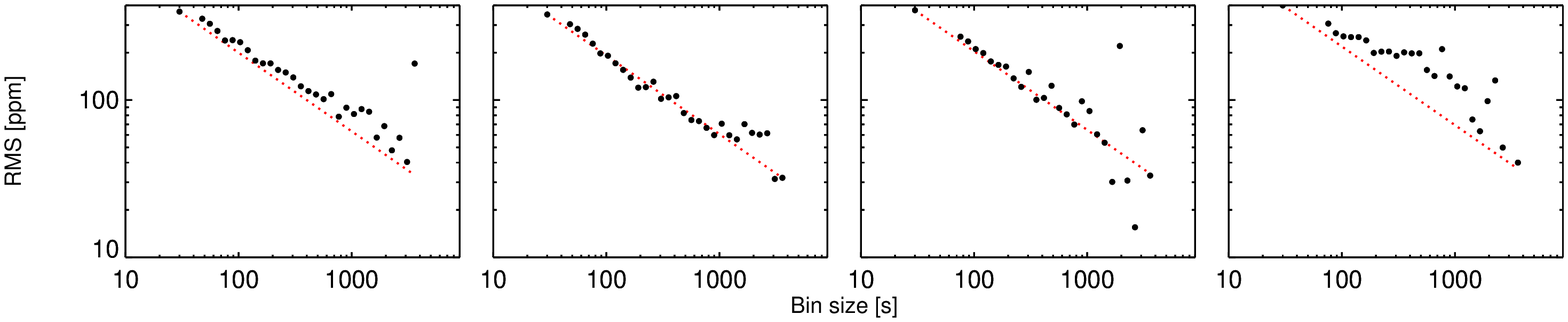}
\caption{\label{fig:trrms13}{\bf 2013 transit  - RMS vs. binning.} Black filled circles indicate the photometric residual RMS for different time bins. Each panel corresponds to each individual transit from 2013. The expected decrease in Poisson noise normalised to an individual bin (30s) precision is shown as a red dotted line. }
\end{figure*}

In this third part of the analysis, we assess whether transit depths vary across our 2011-2013 datasets. This MCMC fit includes the six transit lightcurves only, each of them having an independent planet-star radius ratio $R_{P}/R_{\star}$. To increase the convergence efficiency, we use Gaussian priors on $b$, $P$, $W$ and $T_0$ using the posterior distribution derived from the joint fit in the previous step. Results for this MCMC fit are shown on Table~\ref{tab:tr}. The corrected lightcurves for each transit are shown on Fig.~\ref{fig:tr11} and \ref{fig:tr13} while the corresponding correlated noise behaviour is shown on Fig.~\ref{fig:trrms11} and \ref{fig:trrms13}.

We find that 55\,Cnc\,e's transit depths at 4.5$\mu$m change by 25\% in comparison to the mean value (365 ppm) between 2011 and 2013. Fitting all transits with a straight line yields a reduced chi-squared of $\chi^2_r$ = 1.2, corresponding to an overall variability significance of 1$\sigma$, which we consider being undetected. While most transit depth values are consistent at the 1-$\sigma$ level, only the first 2011 transit appears as an outlier. Converting our transit depths to planetary sizes, we find a minimum planet radius of 1.75$\pm$0.13 Earth radii, a maximum radius of 2.25 $\pm$0.17 Earth radii and an average value of 1.92$\pm$0.08 Earth radii. This value is significantly smaller than the published {\it Spitzer}+MOST combined estimate of  2.17 $\pm$0.10 Earth radii \citep{Gillon:2012}. Alternatively, these marginal transit depth variations could be due to steep variations of the stellar flux, which is however unlikely for this star (see Sect.~\ref{var}).

\begin{table*}
\begin{tabular}{ll}

\hline
Date obs & Transit depth [ppm] \\
\hline
2011-01-06 & 484$\pm$74 \\
2011-06-20 & 287$\pm$52 \\
2013-06-21 & 325$\pm$76 \\
2013-07-03 & 365$\pm$43 \\
2013-07-08 & 406$\pm$78 \\
2013-07-11 & 433$\pm$90 \\
\end{tabular} 

\caption{\label{tab:tr} {\bf 55\,Cnc\,e Spitzer transit depths.} IRAC 4.5$\mu$m transit depths derived from the MCMC simulation detailed in Sect.~\ref{supp:tr}}

\end{table*}

\subsubsection{Occultation depth variations}
\label{supp:occ}

\begin{figure*}
\centering
\includegraphics[scale=0.6]{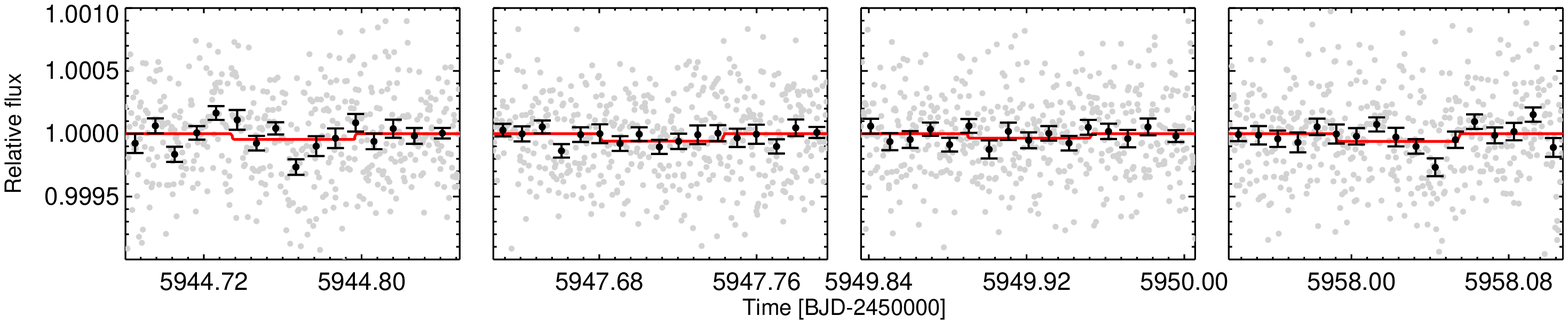}
\caption{\label{fig:occ12}{\bf 2012 occultation lightcurves.} Individual analyses of the Spitzer/IRAC 4.5-$\mu$m 2012 occultation data detrended and normalised following the technique described in the text. Black filled circles are data binned per 15 minutes. The best-fit model resulting from the MCMC fit is superimposed in red for each eclipse.}
\end{figure*}

\begin{figure*}
\centering
\includegraphics[scale=0.6]{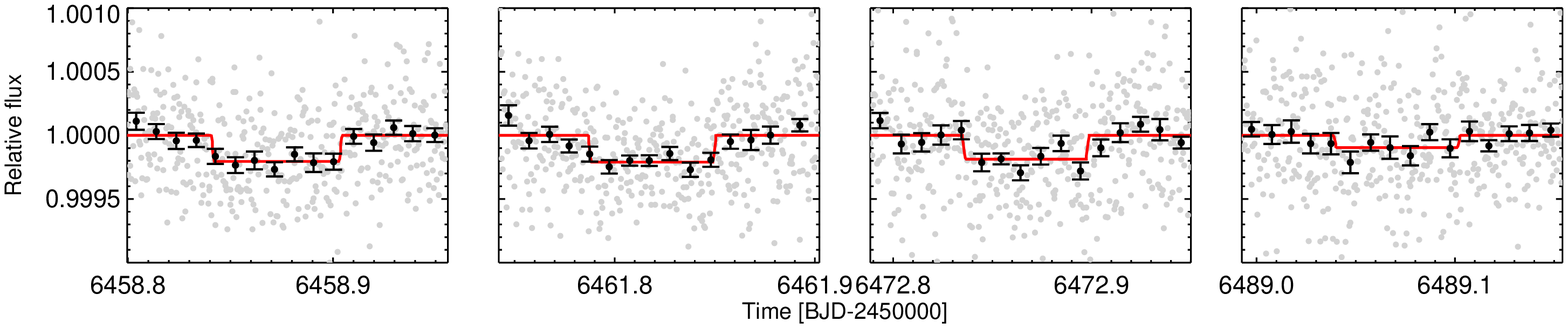}
\caption{\label{fig:occ13}{\bf 2013 occultation lightcurves.} Individual analyses of the Spitzer/IRAC 4.5-$\mu$m 2013 occultation data detrended and normalised following the technique described in the text. Black filled circles are data binned per 15 minutes. The best-fit model resulting from the MCMC fit is superimposed in red for each eclipse.}
\end{figure*}

\begin{figure*}
\centering
\includegraphics[scale=0.6]{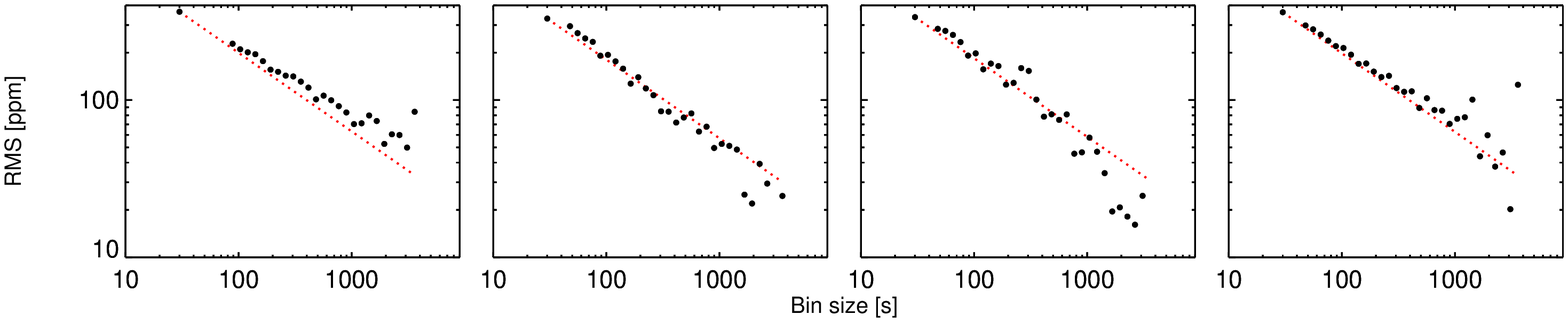}
\caption{\label{fig:occrms12}{\bf 2012 occultation - RMS vs. binning.} Black filled circles indicate the photometric residual RMS for different time bins. Each panel corresponds to each individual occultation from 2012.  The expected decrease in Poisson noise normalised to an individual bin (30s) precision is shown as a red dotted line. }
\end{figure*}

\begin{figure*}
\centering
\includegraphics[scale=0.6]{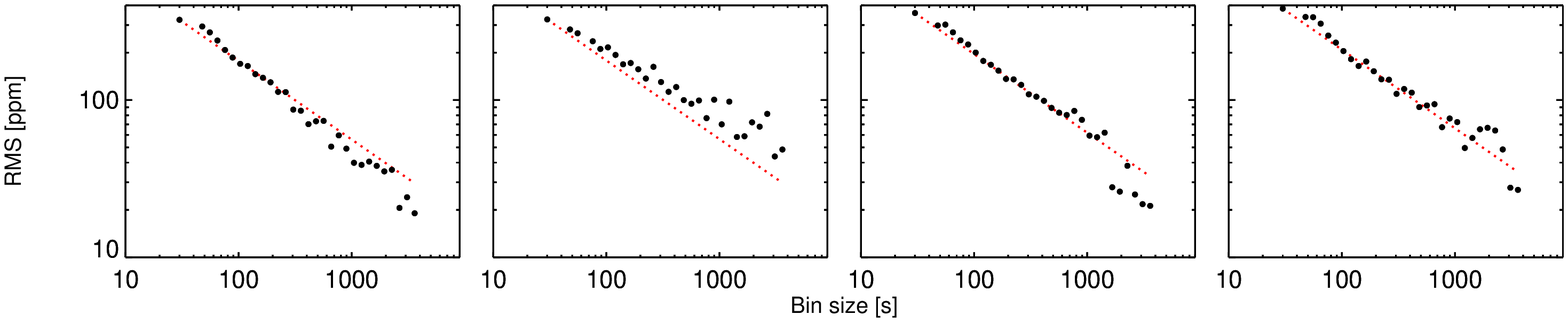}
\caption{\label{fig:occrms13}{\bf 2013 occultation - RMS vs. binning.} Black filled circles indicate the photometric residual RMS for different time bins. Each panel corresponds to each individual occultation from 2013.  The expected decrease in Poisson noise normalised to an individual bin (30s) precision is shown as a red dotted line. }
\end{figure*}

In the following we assess the extent of variability in the planetary thermal emission.
We first perform an MCMC fit that includes all eight occultations observed in 2012 and 2013, each of them having an individual depth $\delta_{occ}$. All other parameters derived in Sect.~\ref{syspar} are included as Gaussian priors in this MCMC fit. Results for this MCMC fit are indicated in Table~\ref{tab:occ}. The corrected lightcurves for each occultation are shown on Fig.~\ref{fig:occ12} and \ref{fig:occ13} while the corresponding correlated noise behaviour is shown on Fig.~\ref{fig:occrms12} and \ref{fig:occrms13}.  Fitting all eight occultations with a straight line yields a mean occultation depth of 83$\pm$14 ppm and a corresponding $\chi^2_r$ = 2.6.

\begin{figure*}
\centering
\begin{subfigure}[b]{0.5\textwidth}
\includegraphics[width=\textwidth]{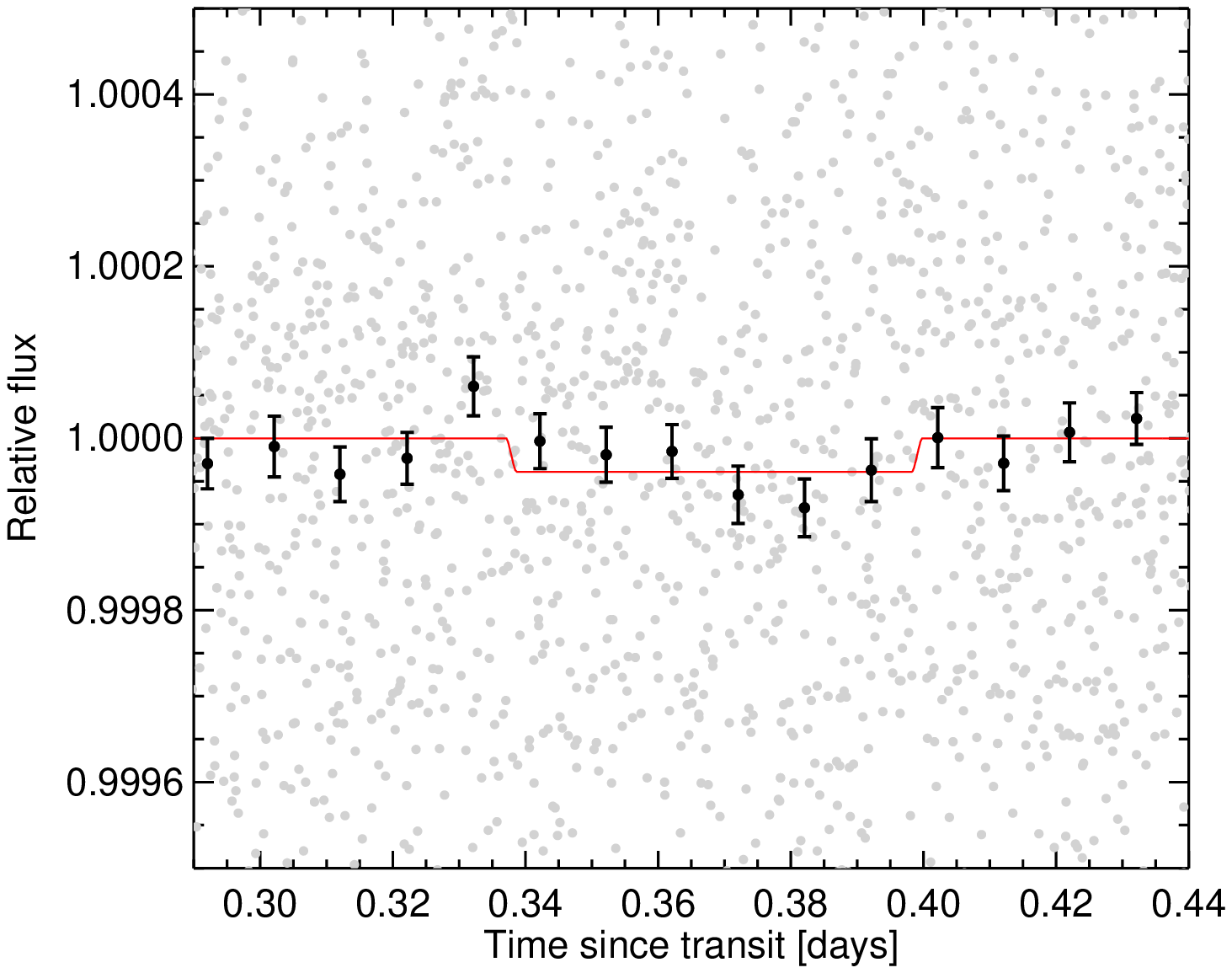}
\end{subfigure}
 
\begin{subfigure}[b]{0.5\textwidth}
\includegraphics[width=\textwidth]{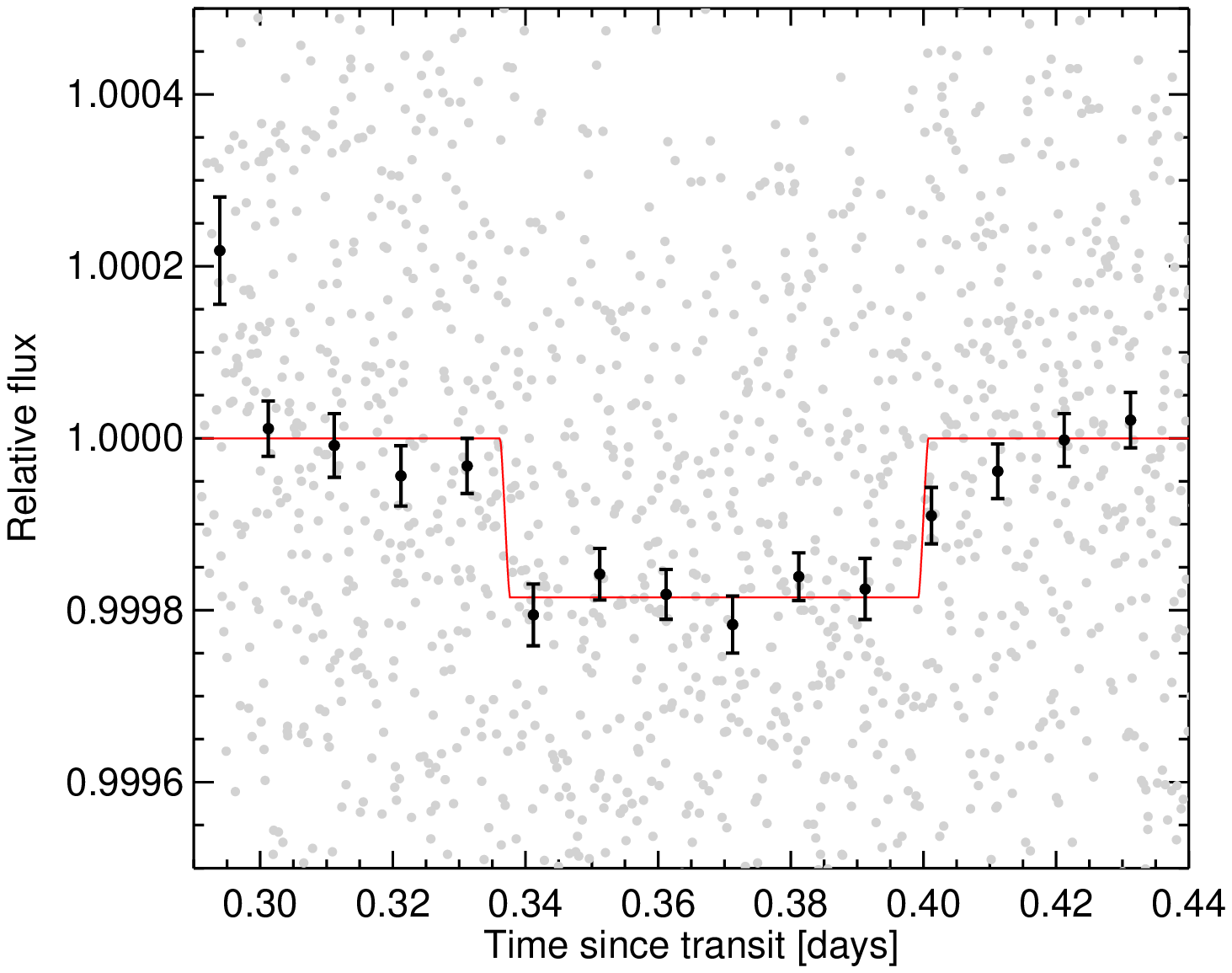}
\end{subfigure}
\caption{\label{fig:lc}{\bf 55\,Cnc\,e Spitzer/IRAC 4.5$\mu$m 2012 vs. 2013 occultations.} The top and bottom lightcurves are the combined occultations for 2012 and 2013 respectively. Each season results from co-adding four occultations. Data are detrended from instrumental systematics and normalised following the technique described in the text. Black datapoints are binned per 15 minutes. Best-fit occultation models are superimposed in red.}
\end{figure*}

\begin{figure*}
\centering
\includegraphics[scale=0.7]{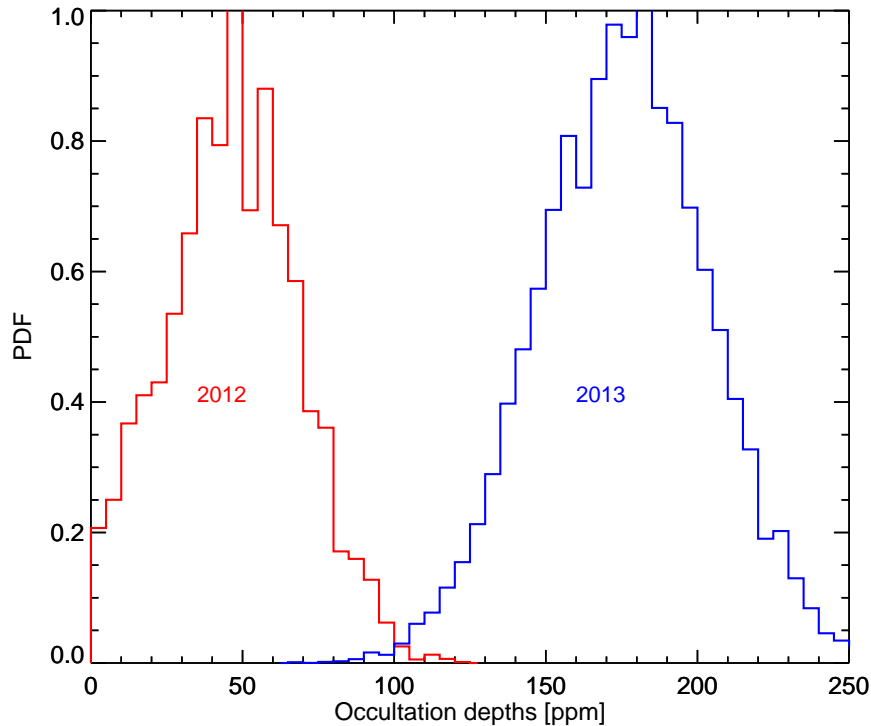}
\caption{\label{fig:histo}{\bf 55\,Cnc\,e average occultation depths $\delta_{occ}$ in 2012 and 2013.} Posterior probability distribution functions for the 2012 (red) and 2013 (blue) mean occultation depths. Both histograms indicate the entire credible region constrained by the {\it Spitzer} data using an MCMC fit.}
\end{figure*}

In a second MCMC fit, we independently fit the occultations by season to investigate a difference in depth between 2012 and 2013. We find that fitting both the 2012 and 2013 occultation depth values with a straight line yields $\chi^2_r$ = 13.6, meaning that constant occultation depths fits the data poorly at the 3.7-$\sigma$ level.  We find that the occultation depths as measured with {\it Spitzer} at 4.5 $\mu$m vary by a factor of $3.7^{+4.1}_{-1.6}$ between 2012 and 2013, from 47$\pm$21 to 176$\pm$28 ppm. The resulting lightcurves for each season are shown on Fig.~\ref{fig:lc} and their posterior distributions on Fig.~\ref{fig:histo}. To interpret the variability in the occultation depth, we use an observed infrared spectrum of 55\,Cnc \citep{Crossfield:2012d} to estimate the corresponding change in the planet's thermal emission. Assuming a mean radius of 1.92 Earth radii, we find that between 2012 and 2013, the planet dayside brightness temperature at 4.5$\mu$m changes from $1365^{+219}_{-257}$\,K to $2528^{+224}_{-229}$\,K. 

\begin{figure*}
\centering
\includegraphics[scale=0.7]{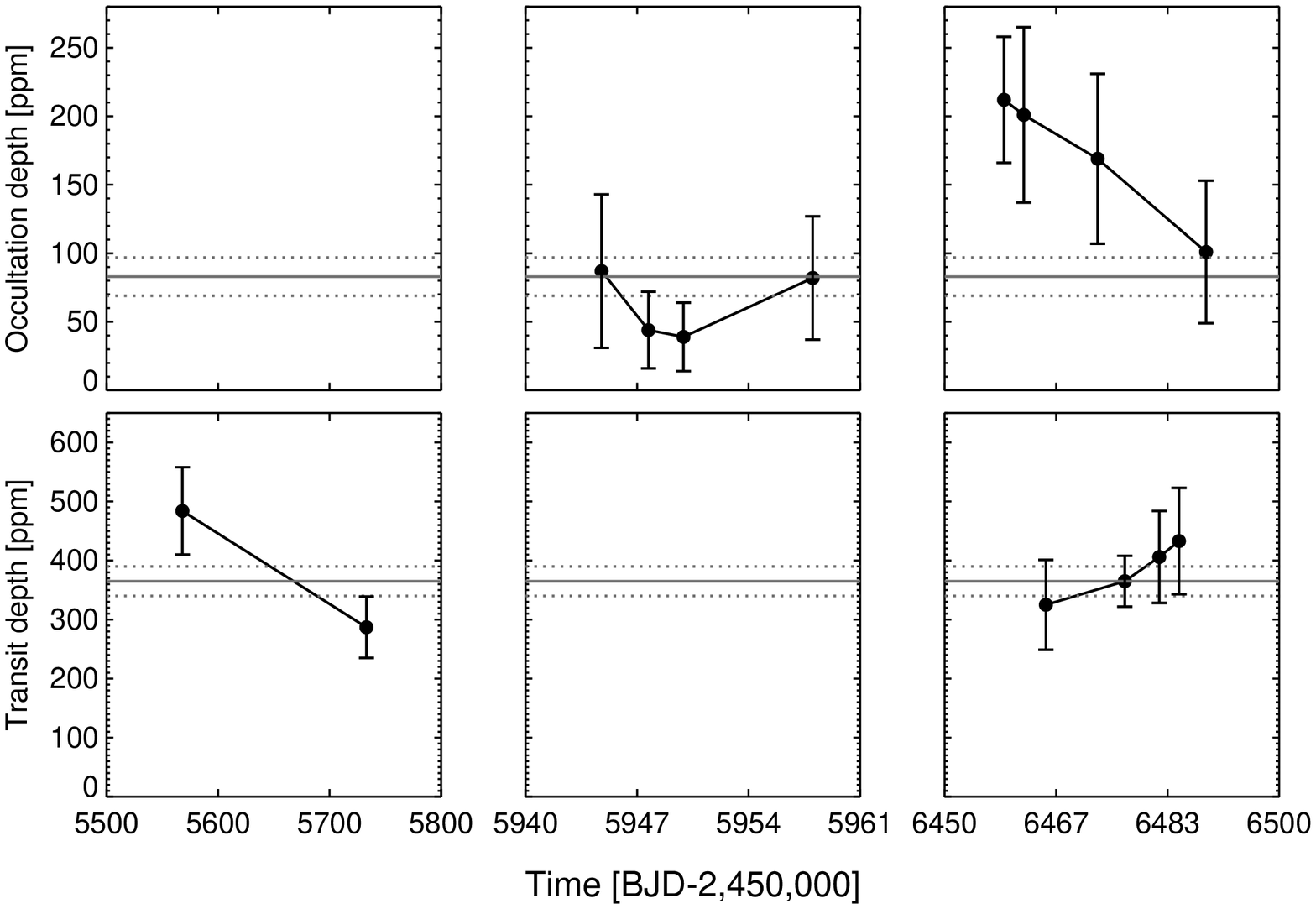}
\caption{\label{fig:depths}{\bf 55\,Cnc\,e Spitzer/IRAC 4.5$\mu$m occultation and transit depths.} Top and bottom panels indicate the occultation and transit depths with time. Left panels are for 2011, middle panels for 2012 and right panels for 2013. The grey solid horizontal lines indicate the mean depths and the dotted lines the 1-$\sigma$ credible intervals.}
\end{figure*}

The time coverage of our combined 2011-2013 dataset (Fig.~\ref{fig:depths}) is too scarce to put firm constraints on the variability period of the emission and apparent size of 55\,Cnc\,e, motivating further monitoring of this planet.

\begin{table*}
\begin{tabular}{lll}

\hline
Date obs & $\delta_{occ}$ [ppm] & $T_B$ [K]\\
\hline
2012-01-18 & 87$\pm$56 & $1767^{+491}_{-592}$\\
2012-01-21 & 44$\pm$28 & $1331^{+292}_{-374}$\\
2012-01-23 & 39$\pm$25 & $1273^{+271}_{-348}$\\
2012-01-31 & 82$\pm$45 & $1720^{+402}_{-470}$\\
2013-06-15 & 212$\pm$46 & $2816^{+358}_{-368}$\\
2013-06-18 & 201$\pm$64 & $2729^{+499}_{-522}$\\
2013-06-29 & 169$\pm$62 & $2472^{+493}_{-524}$\\
2013-07-15 & 101$\pm$52 & $1894^{+446}_{-507}$\\
\end{tabular} 
\caption{\label{tab:occ} {\bf 55\,Cnc\,e Spitzer occultation depths and brightness temperatures.} IRAC 4.5$\mu$m occultation depths derived from the MCMC simulation detailed in Sect.~\ref{supp:occ}. Brightness temperatures are computed using a stellar spectrum, as detailed in the text.}
\end{table*}

\subsection{Discrepancy with the Demory et al. 2012 analysis}

We investigate the discrepancy between our updated 2012 occultation depth, 47$\pm$21 ppm, and our previous one published in \citet{Demory:2012}, 131$\pm$28 ppm, assuming that the different models that we used in the two analyses are at the origin of the discrepancy. Indeed, in our 2012 work, we had not yet noticed the strong dependency of the measured fluxes on the measured widths of the PRF in the $x$ and/or $y$ directions. The improvement brought by the addition of the PRF FWHM in the baseline model is described in detail in \citet{Lanotte:2014} for the case of GJ\,436. In 2012, our modelling of the Spitzer systematics was based uniquely on polynomials of the measured $x$ and $y$ position of the PRF centre. Furthermore, to reach a decent level of correlated noise in the residuals of our fits, we had to add to our models quadratic functions of time and, for three occultations out of four, quadratic functions of the logarithms of time \citep[][Table~1]{Demory:2012}. The resulting complex time-dependent models could have easily ``twisted'' the light curves to improve the fit, affecting the measured occultation depth by a systematic error. To test this hypothesis, we perform several MCMC analyses of the 2012 Spitzer light curves (as obtained with a photometric aperture of 3 pixels). For all these analyses, the transit parameters (depth, impact parameter, ephemeris) are kept under the control of Gaussian priors based on the result of our global analysis presented in Sect.~\ref{syspar}, and assuming a circular orbit.

In our first analysis (T1), we assume a common depth for all four 2012 occultations, and the same baseline models than in the present work (Sect.~\ref{ips}) based on functions of the PRF centre position and widths. In our second test analysis (T2), we assume again a common depth, and we use as baseline models the complex functions used in Demory et al. (2012, Table 1). Our third (T3) and fourth (T4) test analyses are similar, respectively, to T1 and T2, except that we allow each occultation to have a different depth. 

The results of these tests are summarised in Table~\ref{tab:disc1}. By comparing T1 and T2, it can be seen that the models adopted in \citet{Demory:2012} result in a larger and more uncertain occultation depth, to significantly larger scatters of the residuals, and to much larger values of the BIC. The simpler instrumental models used in this present work, based only on measured instrumental parameters (PRF position and width) result in a much better fit to the data {\it and} a reduced and more accurate occultation depth. To confirm this conclusion, we use the four 2013 55\,Cnc\,e occultation light curves. We discard the real  occultations, and inject a fake occultation signal of 100 ppm, assuming a circular orbit for 55\,Cnc\,e, the median of the posteriors of our global analysis for the planetary and stellar parameters (Sect.~\ref{syspar}), and a shift of 0.16d for the inferior conjunction timing. We then perform two global analyses of the four light curves, keeping again the stellar and planetary parameters under the control of Gaussian priors based on the results of our global analysis, and taking into account the same 0.16d shift for the transit timing. In our first analysis, we use our nominal instrumental model based only on low-order (1 or 2) polynomial functions of the $x$- and $y$-positions of the PRF centre and its $x$- and $y$-widths. In our second analysis, we omit the PRF widths terms and add quadratic functions of time and logarithm of time, as in Demory et al. (2012). Table~\ref{tab:disc2} compares the results for both analyses. These results show clearly that our updated instrumental model fit the data better and result in a more accurate measurement of the occultation depth. 

\begin{table*}
\begin{tabular}{lllll}

\hline
Analysis \#       &  Occ. depth  [ppm] & BIC & $\chi^2$ & RMS/15 min [ppm]\\
\hline
T1 & 47$\pm$21 & 2938 & 2596.1 & 64 \\
T2 & 156$\pm$37 & 3583 & 3068.5 & 96 \\
T3 & 101$\pm$55 & 2983.4 & 2592.9 & 64 \\
   &  61$\pm$37 & & &  \\
   &  94$\pm$33 & & &  \\  
   &  152$\pm$46 & &  & \\   
T4 & 220$\pm$64 & 3618.6 & 3056.1 & 93 \\
   &  55$_{-38}^{+56}$ & & &  \\
   &  132$_{-84}^{+120}$ &  & & \\  
   &  217$_{-61}^{+55}$ &  & & \\   
   \end{tabular} 
   \caption{\label{tab:disc1} {\bf 55\,Cnc baseline model test.} The table indicates the results of the different baseline model tests conducted to investigate the discrepancy with the results of Demory et al. 2012 (D12). Analysis T1 shows the fit results from the present analysis, using the same data from D12 and the baseline model presented in this study. T2 uses the same photometry but with the baseline models from D12. T3 and T4 are similar to T1 and T2 respectively but allow each occultation depth to be fit independently.}
\end{table*}
   
\begin{table*}
\begin{tabular}{llll}
\hline
Model        &      BIC  &   RMS/15 min [ppm]   &   Occ. depth [ppm] \\
\hline
p(x,y,w$_x$,w$_y$)  &  2905    &   64   &      97$\pm$22 \\
p(x,y,t,r)    &    3585    &  118     &   63$\pm$34 \\
\end{tabular}
\caption{\label{tab:disc2} {\bf 55\,Cnc\,e occultation injection test.} The table indicate how well our fake 100-ppm deep occultation is recovered by the baseline model used in the present study $p$($x$,$y$,w$_x$,w$_y$) and by the baseline model used in D12 $p$($x$,$y$,$t$,$r$). $x$, $y$ are the PSF positions, $t$ is time, $r$ is the logarithmic ramp and w$_x$,w$_y$ are the width of the PSF along the $x$ and $y$ axes respectively.}
\end{table*}

\subsection{Stellar Variability}
\label{var}

55\,Cnc benefits from a 11-yr extensive photometric monitoring at visible wavelengths obtained with the Automatic Photometric Telescope at Fairborn Observatory. 55\,Cnc has been found to be a quiescent star showing on rare occasions variability at the 6 milli-magnitude level, corresponding to a $<$1\% coverage in star spots \citep{Fischer:2008}. When detected, this variability has a periodicity of about 40 days, corresponding to the rotational period of the star. The variation in stellar surface brightness along the planet path on the stellar disk is directly proportional to the variation in transit depth \citep{Mandel:2002,Agol:2010}. Conservatively assuming an average stellar surface brightness variation of 1\% at 4.5$\mu$m cannot be causing the transit and occultation variability of 20 and $\sim$300\% respectively observed for 55\,Cnc\,e. We assume that the assessment of variability made in the past on the 11-yr baseline remains valid for our observations. We do not have optical data that are contemporaneous with our {\it Spitzer} data. The extensive photometric monitoring of the host star, underlining the quiescent nature of 55\,Cnc, confirms that the source of the observed variability originates from the planet or its immediate environment.

\section{Discussion}
\label{disc}
The observed variability in thermal emission and transit depths of the planet seemingly requires an episodic source of 4.5-$\mu$m opacity in excess of the steady state atmospheric opacity. Thermal emission from the dayside atmosphere varies by $\sim$300\%, detected at over 4-$\sigma$, leading to  temperatures varying between $\sim$1300-2800 K assuming a stable photosphere in the 4.5$\mu$m bandpass. Similarly, the amplitude of the variation in the planetary radius in our 2013 observations at 4.5 $\mu$m ranges between 0.2-0.8 Earth radii, corresponding to a radius enhancement of 11-49 \%, albeit detected only at a 1-2 $\sigma$ level.  The amplitudes of these variations require excess absorption equivalent to ten(s) of scale heights of optically thick material absorbing in the IRAC 4.5$\mu$m band. Local changes in the chemical composition or temperature profiles of the observable atmosphere are unlikely to contribute such large and variable opacity on the time-scales of our observations which, nevertheless, need to be confirmed by hydrodynamical simulations of the super-Earth atmosphere \citep[e.g.][]{Zalucha:2013,Carone:2014,Kataria:2015}.

Recent studies motivate the question of whether volcanic outgassing from the planetary surface might contribute to the observed variability. Given the high equilibrium temperature of the planet ($\gtrsim 2000$K) the planetary lithosphere is likely weakened, if not molten, for most mineral compositions thereby leading to magma oceans and potential volcanic activity, particularly on the irradiated day side \citep{Gelman:2011,Elkins-Tanton:2012}. Recent inferences of disintegrating exo-mercuries in close-in orbits  have also used volcanic outgassing as a possible means to inject dust into the planetary atmosphere before being carried away by thermal winds \citep[e.g.][]{Rappaport:2012,Rappaport:2014a}.  However, whereas extremely small planets (nearly Mercury-size) subject to intense irradiation can undergo substantial mass loss through thermal winds, super-Earths are unlikely to undergo such mass-loss due to their significantly deeper potential wells \citep{Perez-Becker:2013,Rappaport:2014a}. Thus, ejecta from volcanic eruptions on even the most irradiated super-Earths such as 55\,Cnc\,e are unlikely to escape the planet and would instead display plume behaviour characteristic to the solar system \citep{Spencer:1997,Spencer:2007,McEwen:1998,Sanchez-Lavega:2015}. The extent and dynamics of the plumes if large enough can cause temporal variations in the planetary sizes and brightness temperatures and hence in the transit and occultation depths.

Volcanic plumes may be able to explain the peak-to-trough variation in thermal emission that we observe from 55\,Cnc\,e. Ascending plumes could raise the planetary photosphere in the 4.5 $\mu$m to lower pressures higher up in the atmosphere where the temperatures are lower, thereby leading to lower thermal emission as observed in Fig.~\ref{fig:depths}. Note that the temperature structure is governed largely by the incoming visible radiation and visible opacity in the atmosphere, whereas the infrared photosphere is governed primarily by the infrared opacity. Therefore, for the temperature profile to remain cooler in the upper-atmosphere the visible opacity in the atmosphere and in the rising plume would need to be weaker compared to the infrared opacity. The detailed hydrodynamics of a potential volcanic plume is outside the scope of the current work, but we can nevertheless assess its plausibility using simple arguments. 

\begin{figure*}
\centering
\includegraphics[scale=0.9]{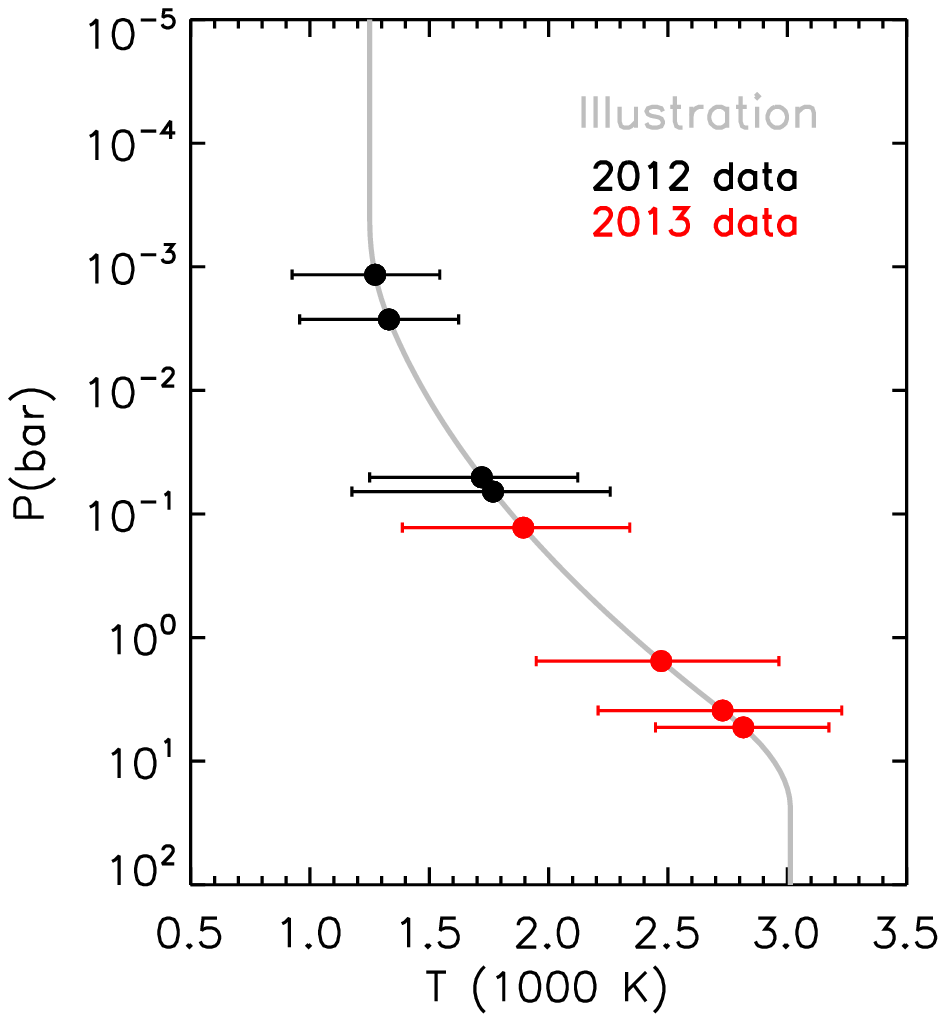}
\caption{\label{fig:p-t}{\bf 55\,Cnc\,e arbitrary pressure-temperature profile.} Schematic representation of the observed brightness temperatures in the IRAC 4.5$\mu$m relative to an illustrative temperature profile of the dayside atmosphere of 55\,Cnc\,e. The data show the measured brightness temperatures along with their uncertainties. The solid gray curve shows a schematic temperature profile as discussed in Section~\ref{disc}.}
\end{figure*}

The observed IRAC 4.5-$\micron$ brightness temperature (T$_{\rm B}$) of the planet varies widely between T$_{\rm min} = 1273^{+271}_{-348}$K and T$_{\rm max} = 2816^{+358}_{-368}$K. If we assume that the T$_{\rm min}$ and T$_{\rm max}$ observed do indeed represent the maximum and minimum temperatures in the atmosphere, they provide a constraint on the temperature profile of the atmosphere. For extremely irradiated atmospheres the temperature profile approaches an isotherm in the two extremes of optical depth; i.e. a skin temperature at low optical depths and the diffusion approximation at high optical depths \citep[see e.g.][]{Madhusudhan:2009c,Spiegel:2010b,Guillot:2010c,Heng:2012a}. Therefore, T$_{\rm B, min}$ and T$_{\rm B, max}$ could represent the two extreme isotherms. The pressures corresponding to the two limits depend on the atmospheric composition, but generally lie between P$_{\rm max} $ of $\sim$ 1-100 bar and P$_{\rm min}$ of $\sim$10$^{-3}$ bar in models of irradiated atmospheres. Fig.~\ref{fig:p-t} shows an illustrative model of a $P$-$T$ profile between (T$_{\rm min}$, P$_{\rm min}$) and (T$_{\rm max}$, P$_{\rm max}$). Note that the curvature of the $T$-$P$ profile and hence the vertical location of the intermediate points are arbitrary in this toy model. The more important matter is that the maximum extent of the plumes can be encompassed within $\sim$10 scale heights of the atmosphere, implying $\sim 500$ km in an H$_2$O-rich atmosphere, and  200 or 300 km for a CO$_2$ or CO atmosphere, respectively. Such plume heights are comparable to those of the largest plumes in the solar system as discussed below \citep{Spencer:1997,McEwen:1998}. However, depending on their aspect ratios and locations relative to the sub-stellar point multiple plumes may be required to influence the disk integrated thermal emission observed.

The chemical composition of the outgassed material is unconstrained by our current observations and would depend on the interior composition of the planet. However, any candidate plume material is required to have significant opacity in the IRAC 4.5 $\mu$m bandpass. Gases such as CO$_2$, CO, and HCN could provide such opacity \citep[e.g.][]{Madhusudhan:2012}. Using internal structure models of super-Earths \citep{Madhusudhan:2012a}, we find that the mass and bulk radius of the planet, given by the minimum radius observed, are consistent with an Earth-like interior composition of the planet, i.e. composed of an Iron core (30\% by mass) overlaid by a silicate mantle and crust. Previous studies which used a larger radius of the planet required a thick water envelope \citep{Demory:2011,Gillon:2012, Winn:2011a} or a carbon-rich interior \citep{Madhusudhan:2012a}, neither of which are now required but cannot be ruled out either. However, under the assumption of an Earth-like interior, which is also the case for Io, the outgassed products could comprise of gaseous CO$_2$ and sulfur compounds \citep{Moses:2002,Schaefer:2005} along with particulate matter composed of silicates \citep{Spencer:1997,McEwen:1998}. The steady state composition of the atmosphere can be equally diverse \citep{Miguel:2011,Schaefer:2011,Hu:2014}, although the stability of a thick ambient atmosphere is uncertain \citep{Castan:2011,Heng:2012c}. While gaseous products could contribute significant absorption in the IR, such as the strong CO or CO$_2$ or HCN bands in the 4.5 $\mu$m Spitzer IRAC bandpass in which our observations were made, gaseous and particulate Rayleigh scattering and particulate Mie scattering can cause substantial opacity in the visible and IR  wavelengths. New high-precision high-cadence spectroscopic observations over a wide spectral range, e.g. with {\it HST} and ground-based facilities at present and with {\it JWST} in the future, will be necessary to characterise the chemical composition of the plume material. 

On the other hand, if volcanic plumes occur along the day-night terminator of the planet they could also cause variations in the transit depths. We detect only marginal variation (1-$\sigma$) in the transit depths. Nevertheless, we briefly assess their potential implications in this context. The fractional change in the transit-depth due to  plumes in the vicinity of the day-night terminator is given by $\delta \sim (\sum_{i=1}^{N} \epsilon_{\rm i} H_{\rm i}^2)/A_{0}$, where $A_{0}$ is the apparent area of the planet disk in quiescent phase, $N$ is the number of plumes, $H_{\rm i}$ is the height of plume i and $\epsilon_{\rm i}$ is the aspect ratio of plume i. Considering all the plumes to be of similar extent and an aspect ratio of unity \citep[e.g.][]{Spencer:1997}, the average plume height required to explain the variability is given by $H \sim \sqrt{A_{0}\delta/N}$. For plumes distributed over the entire terminator, the disk-averaged plume height is given by $H_{\rm av} \sim \Delta R_p$, where $\Delta R_p$ is the amplitude of the apparent change in the planetary radius, which implies $H_{\rm av} \sim 1300-5100$ km $\sim 0.1- 0.4$ $R_p$. On the other extreme, if only one plume is involved then $H \sim 1 - 2.2$ $\times 10^4$ km $\sim 1 - 2$ $R_p$. Thus, depending on the number of plumes, the plume height required to explain the maximum variability in transit depths can vary between $H \sim$ 0.1 - 2 $R_p$. The required plume heights are generally larger than those found in the solar system, though not implausible. The highest plumes in the solar system are typically seen in bodies with low gravities, such as Io and Mars. The highest volcanic plumes on Jupiter's moon Io (radius, $R_{\rm Io} = 1821$ km) have heights of $\sim$ 300-500 km or 0.16-0.27 $R_{\rm Io}$ \citep{Spencer:1997,McEwen:1998}. Recently, the highest plume on Mars was reported at the day-night terminator with a height of $\sim$250 km (0.07 $R_{\rm Mars}$) and a width of 500-1000 km, i.e with  aspect ratio over 2:1 \citep{Sanchez-Lavega:2015}. The source of the Martian plume is currently unknown though apparently non-volcanic. 

The required plume heights are in principle attainable given the planet's orbital parameters. The Hill radius of the planet is given by $R_{\rm Hill}  = a (M_p/3 M_s)^{1/3}$ where $a$ is the orbital separation and $M_p$ and $M_s$ are the masses of the planet and star, respectively. The required plume heights are well within the Hill radius of 55 Cnc e, $R_{\rm Hill} \sim 3.5 R_p$, implying that the ejecta is unlikely to escape the system and will fall back to the surface. Furthermore, following \citet{Rappaport:2012}, the sound speed ($v_s$) at the surface of the planet is $\sim$1 km/s which is significantly lower than the escape velocity ($v_{esc}$) of 24 km/s at the planetary surface thereby preventing a Parker-type wind to carry the ejecta beyond the Hill sphere. On the other hand, the velocity of the ejecta required to sustain $H_{\rm av} \sim 1300-5100$ km is between 8-17 km/s, which is below the escape velocity of the planet (24 km/s) and is significantly lower than the maximal ejection velocities of $\sim$300 km/s observed on Io. 

Alternately, the data may also be potentially explained by the presence of an azimuthally inhomogeneous circumstellar torus made of ions and charged dust particles similar to Io \citep{Kruger:2003} that could contribute a variable gray opacity along the line-of-sight through diffusion of stellar light. While CO gas in the torus could provide the required opacity in the 4.5 $\mu$m bandpass, the grey behaviour of dust could provide the same opacity both in the visible and infrared.
Io's cold torus rotates with the orbital period of Jupiter \citep{Belcher:1987}. Assuming the same scenario holds for 55\,Cnc\,e would mean that the torus rotates with the same period as the host star ($\sim$42 days), which would cause a modulation of the transit and occultation depths over similar timescales.
For such a circumstellar torus, the optical depth of the time-varying torus segment occulting the stellar disk governs the observed stellar radius and hence the transit depth; for the same planetary radius, a denser segment causes a deeper transit compared to a lighter segment. Thus, a denser segment gradually coming into view could cause the observed increase in transit depth. Similarly for the occultation depth, increasing the optical depth of the segment could result in a decrease in the occultation depth. Volcanism could provide a source for the torus' material akin to Io \citep{Kruger:2003}. Future observational and theoretical studies are needed to explore this alternative. Constraining the nature of the material surrounding 55\,Cnc\,e will be a direct probe of the planet surface composition and may also constrain the formation pathway of hundreds of similar planets that have been found so far.

\section*{Acknowledgments}

We thank Drake Deming for having performed an independent check of the occultation variability and depths in the {\it Spitzer} 4.5-$\mu$m data using his pixel-level decorrelation technique. B.-O.D. thanks Kevin Heng and Caroline Dorn for discussions. We thank the Spitzer Science Center staff for their assistance in the planning and executing of these observations. We thank the anonymous referee for an helpful review. This work is based in part on observations made with the Spitzer Space Telescope, which is operated by the Jet Propulsion Laboratory, California Institute of Technology under a contract with NASA. Support for this work was provided by NASA through an award issued by JPL/Caltech. M.G. is Research Associate at the Belgian Funds for Scientific Research (F.R.S-FNRS). This research made use of the IDL Astronomical Library and the IDL-Coyote Graphics Library.

\footnotesize{

\bsp
\label{lastpage}

\end{document}